\documentclass[singlecolumn]{aastex631}
\usepackage{blindtext}
\usepackage[caption=false]{subfig}

\usepackage{rotating}
\usepackage{amsmath}
\usepackage{graphicx}
\usepackage{afterpage}

\begin{document}
 
\title{Impact of Changing Stellar and Planetary Magnetic Fields on (Exo)planetary Environments and Atmospheric Mass Loss}


\author[0000-0002-8920-1495]{Sakshi Gupta}
\affiliation{Center of Excellence in Space Sciences India, Indian Institute of Science Education and Research Kolkata, Mohanpur 741246, India}
\affiliation{Department of Physical Sciences, Indian Institute of Science Education and Research Kolkata, Mohanpur 741246, India}

\author[0000-0001-9033-8926
]{Arnab Basak}
\affiliation{Center of Excellence in Space Sciences India, Indian Institute of Science Education and Research Kolkata, Mohanpur 741246, India}

\author[0000-0001-5205-2302]{Dibyendu Nandy}
\affiliation{Center of Excellence in Space Sciences India, Indian Institute of Science Education and Research Kolkata, Mohanpur 741246, India}
\affiliation{Department of Physical Sciences, Indian Institute of Science Education and Research Kolkata, Mohanpur 741246, India}

\begin{abstract}

The magnetic activity of a star -- which modulates the stellar wind outflow -- shapes the immediate environments of orbiting (exo)planets and induces atmospheric loss thereby impacting their habitability. We perform a detailed parameter space study using three dimensional magnetohydrodynamic simulations to understand the effect of changing stellar wind magnetic field and planetary magnetic field strengths on planetary magnetospheric topology and atmospheric losses. It is observed that the relative strengths of stellar and planetary magnetic fields play a significant role in determining the steady state magnetospheric configuration and atmospheric erosion. When the stellar field is strengthened or the planetary field is weakened, stellar magnetic field accumulation occurs at the day-side of the planet forcing the magnetopause to shift closer to its surface. The magnetotail opens up leading to the formation of Alfv\'{e}n wings in the night-side wake region. We demonstrate how reconnection processes and wind conditions lead to the bifurcation of the magnetotail current sheet. With increasing stellar wind magnetic field strength, the day-side reconnection point approaches the planet thereby enhancing mass loss. We establish an analytic equation relating the atmospheric mass loss rates to stellar and planetary magnetic field strengths which successfully explains the modeled behaviour. Our results are relevant for understanding how the interplay of stellar and planetary magnetism influence (exo)planetary environments and their habitability in star-planet systems with differing relative magnetic field strengths, or in a single star-planet system over the course of their evolution with age.

\end{abstract}


\section{Introduction} \label{sec:intro}
A star interacts with the planets it hosts via magnetized stellar wind which directly impacts their magnetospheric configuration and atmospheric loss. Gravitational (tidal) interactions can be neglected if the planet lies far away (as in the case for Sun-Earth system) and in such a scenario, the stellar wind independently affects the planetary environment without the possibility of the coronal structure of the star getting modified. The evolution of planetary atmospheres and the nature of its interaction with the stellar wind are profoundly affected by changes in their respective magnetic field strengths (\citealt{Cohen_2015,basak_nandy_2021}). Discovery of several exoplanets (\citealt{mayor_queloz_1995,pepe_ehrenreich_meyer_2014,lunine_macintosh_peale_2009}) has sparked interest in exploring the signatures of life-sustaining conditions and understanding  habitability from the perspective of the origin and evolution of planetary atmospheres (\citealt{article3, book,pollack,10.1093/mnras/stx1069}). Habitability -- in the astrophysical context -- depends on the ability of the stellar wind forced planet to hold on to its atmosphere \citep{Alvarado_G_mez_2020} and some other aspects; this work focuses on stellar wind planetary interactions.

The age of the star determines its activity which dictates the properties of the outflowing wind. Observations confirm the varying magnetic activity of the Sun during its evolutionary phases (\citealt{NANDY2007891,vidotto_2021}), as well as variations in field strength of solar system planets and stars outside our solar system (\citealt{Stevenson_1983,kiefer_schad_davies_roth_2017,do_Nascimento_Jr__2016}). Exploration of magnetic fields on exoplanets is a research pursuit of high topical interest (\citealt{Oklop_i__2020,article2}). With the detection of diverse magnetic activity in a number of star-planet systems, the question that naturally arises is -- what is the consequence of magnetic interactions of (exo)planets with their host star? This is a critical question because the answer to this determines whether a planet can retain its atmosphere during its long-term interaction with the harboring star and consequently, its habitability time-scale. We explore how the nature of this interaction is determined by the relative strengths of the stellar wind magnetic field and planetary magnetosphere. Several efforts have been made to understand how stellar winds influence the planetary magnetosphere as well as its atmosphere (\citealt{article,vidotto_2014,10.1093/mnras/staa852,Harbach_2021,nandy_martens_obridko_dash_georgieva_2021}). Variation in stellar magnetic activity (\citealt{nandy_2004,NANDY2007891,brun_2014,vidotto_2021,tripathi,nandy_martens_obridko_dash_georgieva_2021}) introduces variations in stellar radiation \citep{Spina_2020}, stellar wind speed \citep{Finley_2018}, magnetic field strength of plasma winds \citep{article1} and may lead to magnetic storms as well. As a consequence of these interconnected phenomena, the planetary dipolar field is deformed resulting in a magnetospheric structure that may vary from what we often observe on the Earth
\citep{gallet_charbonnel_amard_brun_palacios_mathis_2016}.

In this paper, we perform a thorough parameter space study using numerical modelling for understanding interactions in star-planet systems and the effect of stellar activity evolution on planets with different magnetospheric strengths. The magnetized stellar wind and the planet's inherent magnetic field are projected to have a significant impact on planetary habitability and atmospheric mass loss (\citealt{nandy_2004,Khodachenko,nandy_valio_petit_2017}) . We carry out three dimensional global magnetohydrodynamic (MHD) simulations using a broad range of stellar and planetary magnetic field values for providing a comprehensive picture of the interaction process. Most extrasolar giant planets that have been discovered so far are anticipated to have significant ionospheres. To simplify the global simulation and based on prior studies (\citealt{Koskinen_2010, caII, titan}), we consider the planet to be surrounded by a perfectly conducting plasma atmosphere. We study the impact of increasing stellar wind magnetic field on magnetopause stand-off distance and explore the conditions that lead to the magnetic pile-up in the day-side region. Another significant occurrence we examine is the formation of Alfv\'{e}n wings where the magnetotail opens up when the upstream wind Alfv\'{e}nic Mach number is low and the current sheet length shortens and bifurcates in the night-side region of the planet. We find a criterion for current sheet bifurcation that is consistent with all cases of our study and establish an analytical relation for the variation of mass loss rate with changing stellar and planetary magnetic fields.

This study is relevant for any star-planet system and in those systems in which the magnetic fields of either or both entities vary considerably with time. Our findings are significant for a better understanding of (exo)planetary atmospheres and determining the impact of varying magnetic field strength on habitability.
This paper begins with the model description in section \ref{sec:style} in which we give an overview of the theory and numerical setup employed in this study. The detailed findings are presented in section \ref{sec:floats} followed by the conclusion in section \ref{sec:displaymath}.

\section{MODEL DESCRIPTION} \label{sec:style}
We adapt the Star-Planet Interaction Module (CESSI-SPIM) developed by \cite{Das_2019} for simulating different configurations of stellar wind and planetary magnetospheres. The governing set of resistive MHD equations are given by:

\begin{equation}
  \partial_{t} \rho+\nabla \cdot(\rho \vec{v})=0  
\end{equation}
  
\begin{equation}
   \partial_{t} \vec{v}+(\vec{v} \cdot \nabla) \vec{v}+\frac{1}{4 \pi \rho} \vec{B} \times(\nabla \times \vec{B})+\frac{1}{\rho} \nabla P=\vec{g}
\end{equation}

\begin{equation}
   \partial_{t} E+\nabla \cdot[(E+P) \vec{v}-\vec{B}(\vec{v} \cdot \vec{B})+(\eta \cdot \vec{J}) \times \vec{B}]=\rho \vec{v} \cdot \vec{g}
\end{equation}

\begin{equation}
   \partial_{t} \vec{B}+\nabla \times(\vec{B} \times \vec{v})+\nabla \times(\eta \cdot \vec{J})=0
\end{equation}

where the symbols $\rho$, $v$, $\vec{B}$, $P$, $E$, and $\vec{g}$ denote density, velocity, magnetic field, pressure, total energy density and gravitational acceleration due to the planet respectively. $\vec{J}$ is the current density given by $\nabla \times \vec{B}$, ignoring the displacement current. The expression for total energy density is given by

\begin{equation}
    E = \frac{P}{\gamma-1} + \frac{\rho v^2}{2} + \frac{B^2}{8\pi}
\end{equation}
for an ideal gas equation of state.

The computational domain extends from -80 ${R_p}$ to 200 ${R_p}$ in the $x$-direction, -45 ${R_p}$ to 45 ${R_p}$ in the $y$-direction and -200 ${R_p}$ to 200 ${R_p}$ in the $z$-direction, where ${R_p}$ is the radius of the planet which is located at the origin of the Cartesian box. The region extending from -2 ${R_p}$ to 2 ${R_p}$ is resolved by 12 grids in all three directions i.e. 1 ${R_p}$ is resolved using 3 grids in the planetary vicinity. Keeping in mind our modest computational facility, a combination of stretched and uniform grid types is used in the $x$ and $z$ directions. In the $x$-direction, the regions extending from -80 ${R_p}$ to -2 ${R_p}$ and 2 ${R_p}$ to 10 ${R_p}$ are resolved using 156 and 16 grids respectively i.e. 1 ${R_p}$ is resolved using 2 grids. Grids with stretching ratio 1.005175 are used from 10 ${R_p}$ to 200 ${R_p}$. In the $y$-direction, the regions from -45 ${R_p}$ to -2 ${R_p}$ and from 2 ${R_p}$ to 45 ${R_p}$ are resolved by 86 grids each i.e. 1 ${R_p}$ is resolved using 2 grids. In the $z$-direction, the regions from -22 ${R_p}$ to -2 ${R_p}$ and 2 ${R_p}$ to 22 ${R_p}$ are resolved by 40 grids each i.e. 1 ${R_p}$ is resolved using 2 grids whereas grids with stretching ratio 1.004958 are used for the regions extending from -200 ${R_p}$ to -22 ${R_p}$ and 22 ${R_p}$ to 200 ${R_p}$.

In this study, an Earth-like planet is considered with similar mass, radius and tilt while the stellar wind is assumed to have a speed of 270 km/s corresponding to quiet times.
We perform a set of simulations by varying the magnetic field strength of both the planetary magnetosphere ($0.1  {B}_{{e}}, 0.5 {B}_{{e}}, 1  {B}_{{e}}, 2 {B}_{{e}}\thinspace $and$ \thinspace 5  {B}_{{e}}$) and stellar wind (-0.1 nT, -0.5 nT, -1 nT, -2 nT, -5 nT, -10 nT, -30 nT, -50 nT and -75 nT) with all possible combinations. Here ${B}_{{e}}$ = $3.1\times 10^4$ nT denotes surface equatorial magnetic field strength of the Earth's dipolar field. 

\begin{table}[htbp]
\centering
\caption{Values of physical parameters used in the simulations and their respective notations.}
\label{table:1}
\begin{tabular}{ l c c } 
 \hline
 \hline
 Physical quantity & Notation & Value used\\ [2ex] \hline
 Density in ambient medium  & $\rho_{amb}$ & 1.5 $\times 10^{-23}$  $\rm{g\thinspace cm^{-3}}$ \\ 
 Pressure in ambient medium & $P_{amb}$ & 2.49 $\times 10^{-11}$ $\rm{ dyne\thinspace cm^{-2}}$ \\
 Density of stellar wind & $\rho_{sw}$ & 4 $\rho_{amb}$\\
 Velocity of stellar wind & $v_{sw}$ & 2.7 $\times 10^{7}$ $\rm{ cm\thinspace s^{-1}}$\\
 Adiabatic index & $\gamma$ & 5/3\\
 Planetary mass & $M_{pl}$ & 5.972 $\times 10^{27}$ \rm{g}\\
 Planetary radius & $R_{pl}$ & 6.371 $\times 10^8$ \rm{cm}\\
  Intrinsic planetary magnetic field & $B_{pl}$ & 0.1$B_{{e}}$ - 5$B_{{e}}$ $\thinspace^*$\\
 Stellar wind magnetic field strength & $|B_{sw}|$ & 0.1 nT - 75 nT\\
 Magnetic diffusivity & $\eta$ & 10$^{13}$ $\rm{ cm^{2}\thinspace s^{-1}}$\\
  Magnetospheric tilt angle & $\theta_{pl}$ & 11$^\circ$\\
 \hline
\end{tabular}

*\textit{The symbol B$_e$ represents the magnetic field for the case of the Earth's dipole}.
\end{table}

In order to study the effect of stellar activity evolution on planetary atmospheric loss which directly impacts habitability, we initialize a conducting plasma atmosphere surrounding the planet. The density profile in the vicinity of the planet is defined by
\begin{equation*}
\rho_{{pl}} = 10^6 \rho_{{atm}}  \qquad r\leq {R_p}
\end{equation*}
\begin{equation}
\begin{split}
\rho_{{atm}} (r) =  \rho_{{pl}} + \frac{(\rho_{{amb}} - \rho_{{pl}})}{2} \Big[{\tanh}\Big\{ 9\Big(\frac{r}{R_P} - 2\Big)\Big\}&+1 \Big] \\
{R_P} \leq r \leq {3R_p} \thinspace ,
\end{split}
\end{equation}
where ${\rho_{{pl}}}$ and ${\rho_{{amb}}}$ are densities of the planet and ambient medium respectively and $r$ is the radial distance from the origin. For more details on the justifications for the choice of the above atmospheric profile, interested readers may refer to Sec. 2.1 of \cite{basak_nandy_2021}. The pressure distribution in the atmosphere is evaluated by numerical integration of the equation
\begin{equation}
    \frac{{dP}}{{dr}} = -\rho_{{atm}} (r) g(r)
\end{equation}
Here, the gravitational field is given by $g(r) = -\frac{G M_{{pl}}}{r^2}$ where $M_{{pl}}$ is the planetary mass. The pressure inside the planet is evaluated by extrapolating the value of pressure at planet-ionosphere boundary. The density and pressure in the region ${r > 3 R_{{pl}}}$ is initialized to be equal to that in the ambient medium~(Table~\ref{table:1}).

The input parameters at the stellar wind injection boundary (at $x$= -80 ${R_p}$ in $yz$ plane) are obtained by solving the Rankine-Hugoniot magnetized jump conditions for the given shock velocity. A southward oriented (SIMF) stellar wind is injected with density 
${\rho}_{{sw}}$ = 4 ${\rho}_{{amb}}$. For all other boundary faces of the Cartesian box, force-free outflow boundary condition is implemented. The magnetic fields of the wind and (or) planet are varied in each run and the steady state configuration of the planetary magnetosphere and atmospheric loss are analyzed.

\section{RESULTS AND DISCUSSION} \label{sec:floats}
\begin{figure}[hbtp]
    \centering
    \includegraphics[width=18cm]{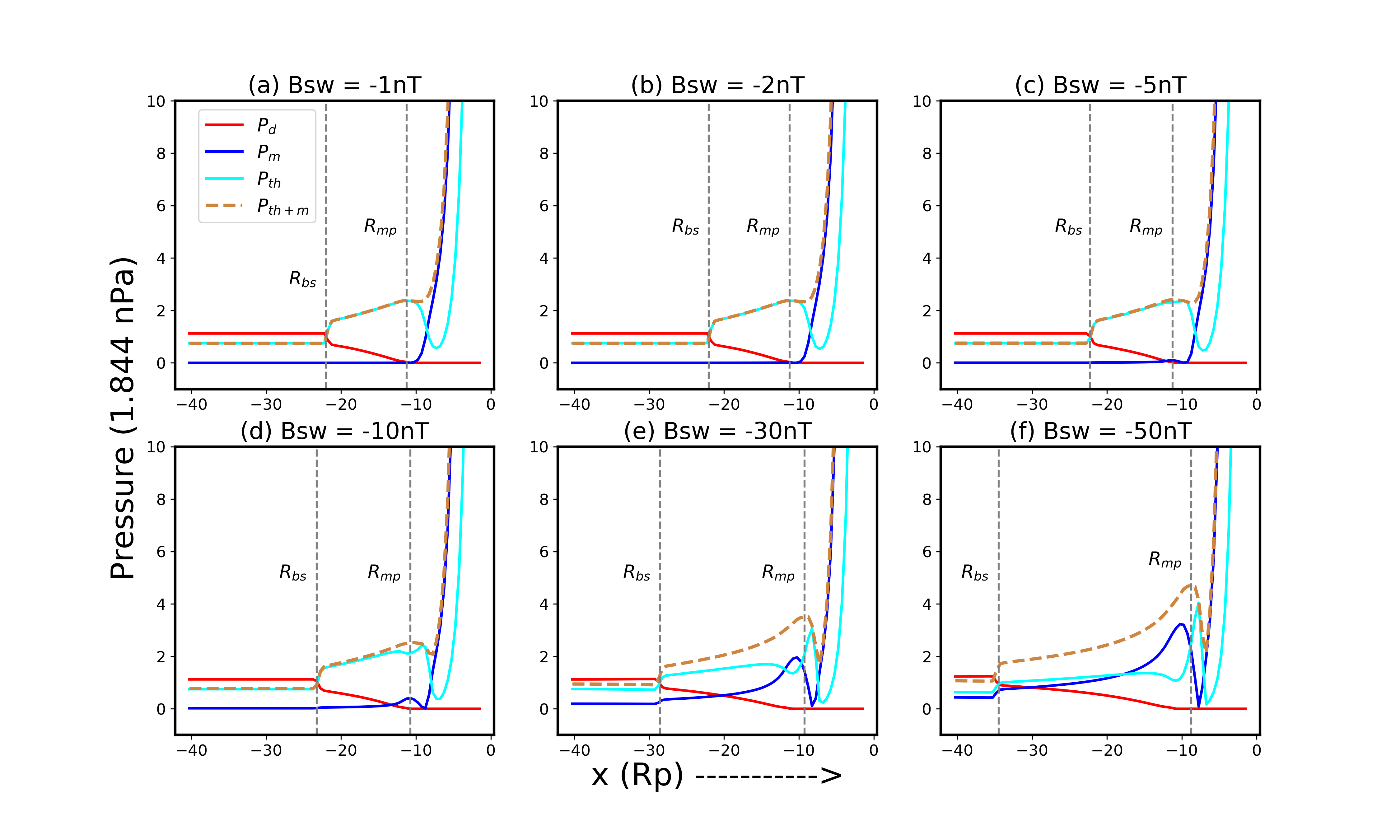}
    \caption{Pressure balance for bowshock and magnetopause: Plots of different pressures in the planetary day-side for magnetospheric field ${B_{p}}$ = ${B_{e}}$ ($3.1\times 10^4$ nT) with varying stellar wind field strength (${B_{sw}}$). The symbols ${P_{th}}$, ${P_{m}}$, ${P_{d}}$, ${P_{th+m}}$ denote thermal, magnetic, dynamic and sum  of thermal and magnetic pressures respectively. The symbols ${R_{bs}}$ and ${R_{mp}}$ denote the positions of the bowshock and the magnetopause stand-off respectively. The value of $R_{bs}$ is obtained by balancing $P_d$ (wind) and $P_{th}$ (magnetosheath) while $R_{mp}$ is evaluated by identifying the location close to the planet where the gradient of $P_{th+m}$ is zero.}%
    \label{fig:magneto_stellar}%
\end{figure}

When the stellar wind impinges upon the planetary magnetosphere, a dynamic pressure balance is achieved at the day-side of the planet resulting in the formation of a magnetopause which inhibits the penetration of stellar plasma and protects the inner atmosphere from erosion. The balance between thermal, magnetic and dynamic pressures determines the shape and location of the magnetopause. An earlier study by \cite{basak_nandy_2021} considers the magnetopause stand-off distance ($R_{{mp}}$) to be the location of balance between the dynamic $P_d$ (incoming wind) and magnetic pressure $P_m$ (near the planet) curves. This approach works well if the stellar wind magnetic field ($B_{{sw}}$) is weak since the magnetic pressure of the incoming wind is very small and may be neglected in the calculations. However, when the stellar magnetic field is relatively stronger, it accumulates outside the magnetopause and the magnetic pressure in the magnetosheath region cannot be ignored any more. Our study shows that the point of intersection between $P_m$ and $P_d$ migrates outward nearer to the bowshock as $|B_{sw}|$ is increased (Figure \ref{fig:magneto_stellar}). For a strongly magnetized wind, it is found that the dynamic pressure $P_d$ is sufficiently high even after this intersection point which indicates that stellar wind plasma is able to penetrate beyond this location and therefore, it cannot be regarded as the magnetopause stand-off distance. Thus, we need a different approach for evaluating the magnetopause stand-off distance for the case of strongly magnetized winds. It has been shown in previous studies (\citealt{schield_1969,shue_chao_2013,lu_wang_kabin_zhao_liu_zhao_li_2015}) that at the magnetopause, the sum of magnetic and thermal pressures (denoted by $P_{{th+m}}$ henceforth) on the planetary side balances the incoming stellar wind (see also \citealt{shue_chao_2013}). These considerations result in the relation

\begin{equation}
    \Big[P_{th} + P_m\Big]_{\rm{planet-side}} =  \Big[P_{th} + P_{m}\Big]_{\rm{star-side}\ .} 
\end{equation}
The distance from the planet where the $P_{{th+m}}$ curve peaks and its gradient becomes zero along the subsolar line is then considered to be the magnetopause stand-off distance.
The intersection of $P_{th}$ and $P_{d}$ curves gives the location of bow shock stand-off distance (\citealt{Liu_Powell_1999, Ma_2004, Holmberg_2019}). We cross validate the estimatates of bow shock stand-off distance by studying the spatial variation of  the Alfvén Mach number; these are in good agreement with the results obtained from our pressure balance analysis.

Figure \ref{fig:magneto_stellar} shows the pressure balance for obtaining bow shock distance (${R}_{{bs}}$) and the magnetopause stand-off distance (${R}_{{mp}}$) for the case of planetary magnetosphere $B_p = B_e$ ($B_e=3.1\times 10^4$ nT) with varying stellar wind magnetic field. Note that only southward interplanetary magnetic field (SIMF) is considered for this study. The total thermal pressure ($P_{{th}}$) is obtained as an output of PLUTO code \citep{Mignone_2007} while dynamic pressure of stellar wind ($P_{d}$ = $\frac{{\rho_{sw} v^2}}{2}$) and magnetic pressure ($P_{m}$ = $\frac{{B^2}}{8{\pi}}$) are evaluated from evolving simulation variables. As the stellar wind magnetic field strength is increased, the magnetopause moves closer to the planet and the bow shock moves further away. For weakly magnetized stellar wind as shown in Figure \ref{fig:magneto_stellar}(a)-(c), the stellar field accumulation is not significant and the intersection of $P_d$ and $P_m$ gives the $R_{mp}$ location as depicted in \cite{basak_nandy_2021}. As $|B_{{sw}}|$ increases further, it is evident that stellar wind magnetic pressure builds up significantly in front of the planet (Figure \ref{fig:magneto_stellar} (d)-(f)). With the increase in $|B_{{sw}}|$, both thermal and magnetic pressure increase in the magnetosheath region initially up to some distance. However, the enhancement in magnetic pressure gets prominent gradually  due to the accumulation of stellar wind magnetic field outside the magnetopause. 

Just after the magnetopause, magnetic pressure drops to nearly zero representing the day-side reconnection $X$ point. Thermal pressure rapidly increases near the magnetopause due to energy conversion in the magnetic reconnection region.  (see e.g. \citealt{lu_wang_kabin_zhao_liu_zhao_li_2015}) 

Pressure balance for different planetary magnetic field strength ($B_{{p}}$) keeping the stellar wind magnetic field constant reveals that with decreasing $B_{{p}}$ both bow shock and magnetopause move towards the planet \citep{basak_nandy_2021}. This implies that for lower values of $B_{{p}}$, stellar wind is able to penetrate closer to the planet resulting in greater atmospheric loss. We find significant increment of magnetic pressure in the magnetosheath region for the case of weaker planetary magnetic field as compared to stronger field cases. Thus, either strengthening the stellar wind magnetic field or weakening the planetary magnetic field leads to magnetic field accumulation at the day side of the planet. Hence, the ratio of the intrinsic planetary magnetic field and stellar wind magnetic fields is an essential factor which determines the dynamics of the star-planet interaction. This is in keeping with expectations.

\begin{figure}[hbtp]
    \includegraphics[width=18cm]{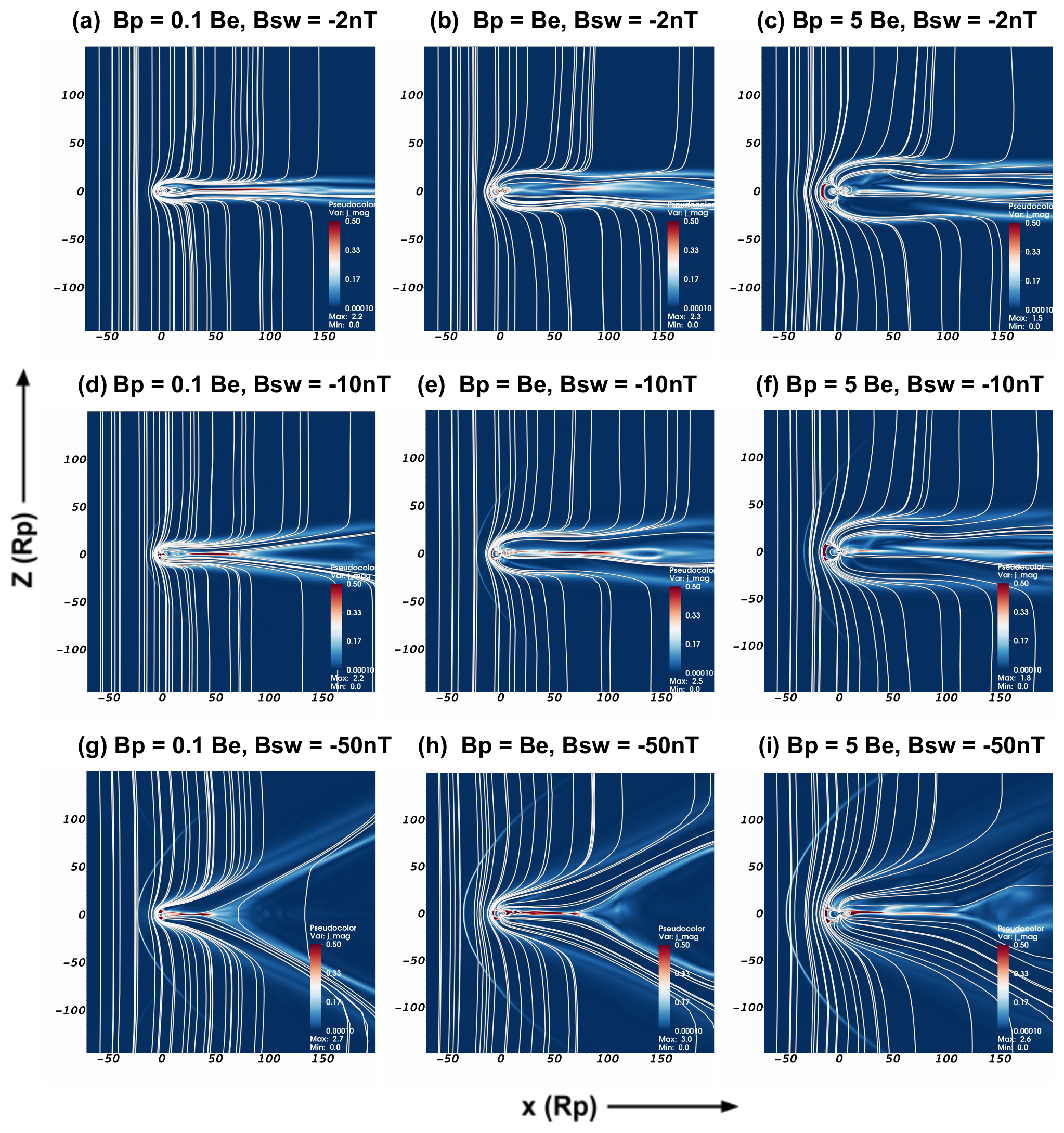}%
    \caption{Steady state planetary magnetospheric configurations for varying stellar and planetary magnetic fields. The background colormap shows current density while magnetic field is plotted using white streamlines in the $y=0$ plane. The planet is located at the origin and the stellar wind flows from left to right. Stellar wind magnetic field increases along each row (top to bottom) while planetary magnetic field strength increases along each column (left to right). Distances are measured in units of the planetary radius.} %
    \label{fig:simulation_result}
\end{figure}

The interaction of plasma winds and planetary magnetosphere results in certain deformation in the planetary dipolar field. This deformation leads to a tear-drop shaped magnetospheric structure that we usually observe for the Earth.
A series of simulation results of interaction of planetary magnetosphere with stellar wind magnetic field is shown in Figure  \ref{fig:simulation_result}. In each subplot, magnetic field lines (white sreamlines) are traced over the current density (colormap) in the $y = 0$ plane with stellar wind plasma flowing from left to right. In plots (a)-(c), the simulations show the formation of extended magnetotail structures for varying planetary magnetic field $B_p=$  0.1 $B_e$, 1 $B_e$, and 5 $B_e$ respectively. Stellar wind magnetic field strength is fixed at $|{B_{sw}}|$ = 2 nT for all cases.

\begin{figure}
    \begin{interactive}{animation}{Alfven_wing.mp4}
        \centering
        \includegraphics[width=16cm]{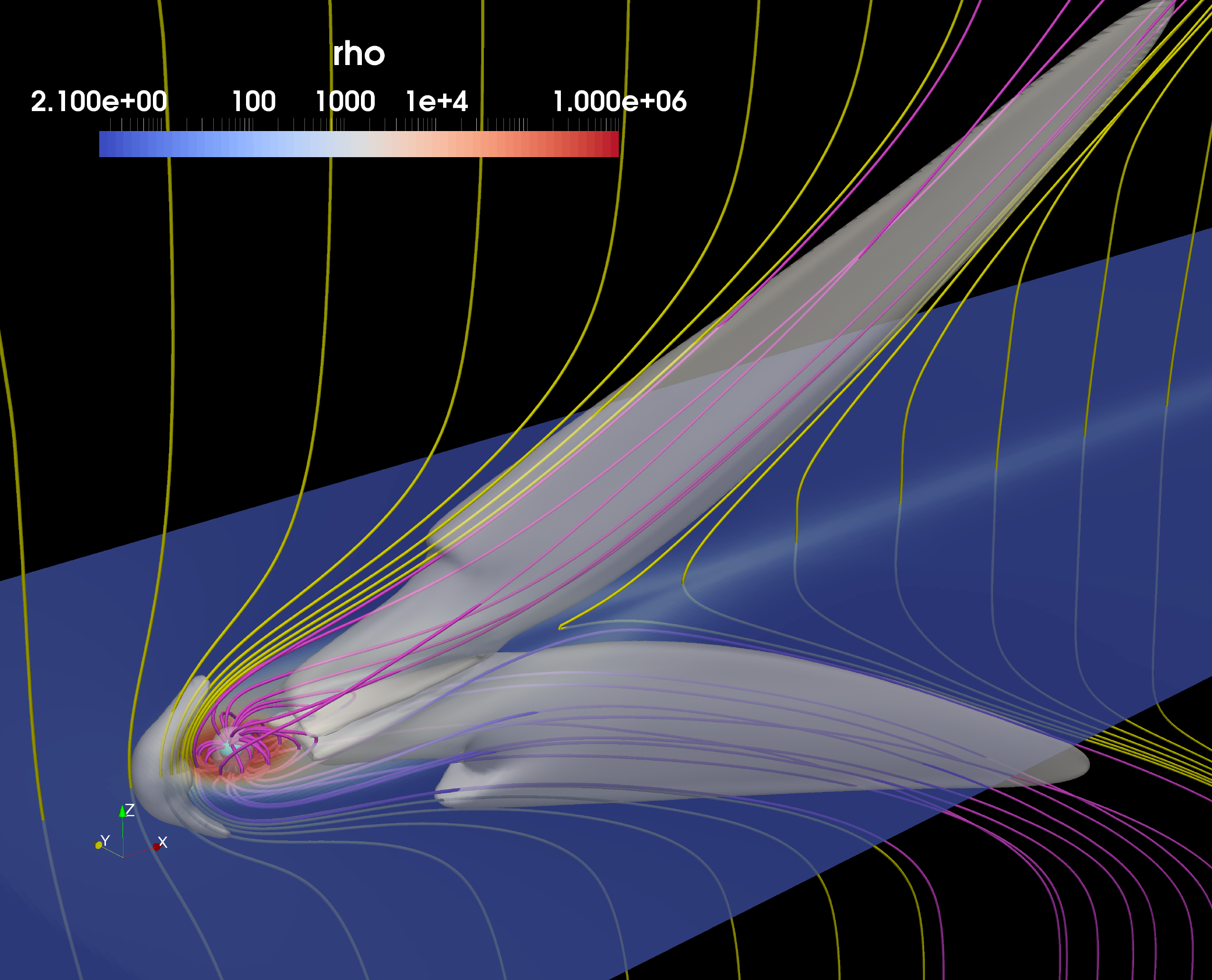}%
    \end{interactive}
    \caption{Steady-state magnetospheric configuration of an (exo)planet with ${B_p} = 0.1 {B_e}$ and ${B_{sw} = -50}$ nT. The white surface encloses regions where Mach number ${M_A}$ $<$ 0.3. Pink and yellow streamlines represent magnetic field lines originating from the planet and pure stellar wind respectively. An animation showing the temporal evolution of this magnetospheric configuration is available online, wherein, the background colors represent the magnitude of current density ($J_{mag}$).}
    \label{fig:Alfven_wings}
\end{figure}

As $|{B_{sw}}|$ is increased to 10 nT [panels (d)-(f)] for the same set of planetary magnetic field values, the magnetotail begins to open up and on further increase of $|{B_{sw}}|$ to 50 nT [panels (g)-(i)] the lobes of the tail open up even further \citep{10.1093/mnras/staa824}. The Alfvén Mach number is an important control parameter that governs the behaviour of plasma embedded in magnetic field. With increasing stellar wind magnetic field strength, the upstream Alfvén Mach number decreases which leads to the appearance of Alfvén wings across the planet \citep {Ridley2007AlfvnWA}. Figure \ref{fig:Alfven_wings} illustrates the sub-Alfvénic plasma interaction with the magnetized planet and the formation of three dimensional Alfvén wings in the planetary magnetotail as shown by the region enclosed by Alfv\'{e}nic Mach number $M_A$ isosurface ($ M_A \leq 0.3$). When stellar wind magnetic field is strong, field lines bend but do not strongly drape the planet. \cite{alfven_wings} have shown the observational evidence of Alfvén wings formation on the Earth during 24 and 25 May 2002. Several moons in our solar system including Lunar wake \citep{zhang_khurana_2016}, Ganymede, Io, Europa, and Enceladus have been observed to form Alfvén wings. \cite{refId0} have found signatures of sub-Alfvénic plasma interactions in several star-(exo)planet systems. With decreasing upstream Alfvén Mach number, wings form at larger angles from the equatorial plane and the length of the current sheet becomes shorter i.e., reconnection takes place closer to the planet in the magnetotail region as shown in column-wise panels of Figure \ref{fig:simulation_result}. The row-wise panels of Figure \ref{fig:simulation_result} show that for a given stellar wind magnetic field strength, Alfvén wings are more prevalent in case of weaker planetary magnetic field. Here, we present only a few simulation results to show how the magnetotail opens up. A similar pattern is found for other simulations that involve all possible combinations of parameters (described in Table \ref{table:1}).  \cite{vernisse_riousset_motschmann_glassmeier_2017} have shown that interactions which involve either a weakly magnetized obstacle or sub-Alfvénic upstream plasma velocities or both can lead to the dominance of an Alfvén wing structure -- and this is corroborated by our simulations.

\begin{figure}[hbtp]
    \centering
    \includegraphics[width=20cm]{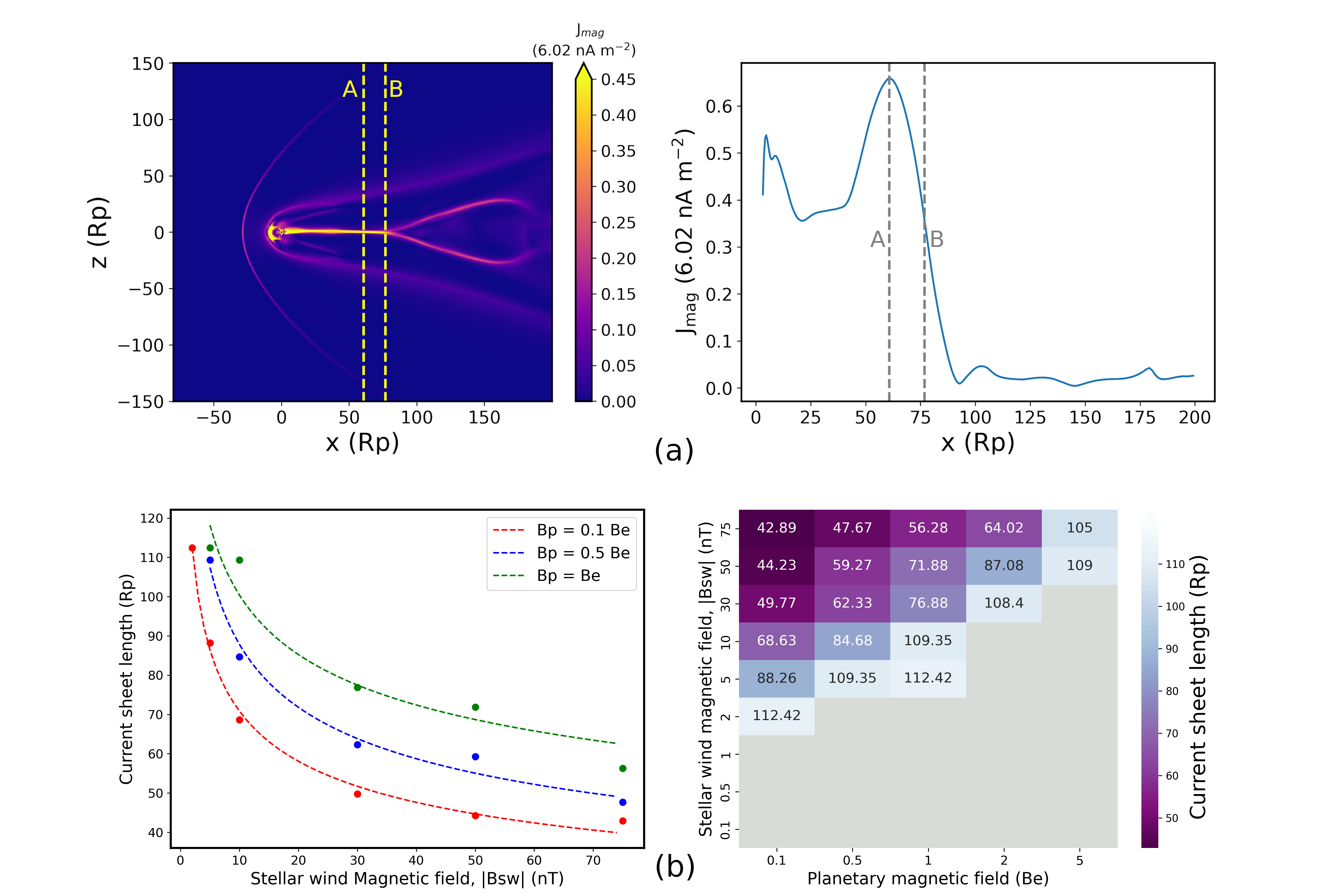}%
    \caption{Current sheet bifurcation due to magnetotail opening: (a) Plot of current density in the $y=0$ plane (left) and corresponding variation of its magnitude ($J_{{mag}}$) along the subsolar line (right) for $B_p$ = $B_e$ and $B_{sw}$ = -30 nT. Line A denotes the location of maximum $J_{{mag}}$ while line B marks the bifurcation point of the current sheet. (b) Left: Variation of  current sheet length with stellar wind magnetic field for three different cases of planetary magnetic field strength. Right: Illustration of current sheet length variation with stellar and planetary magnetic field strength. The light grey region represents cases where the current sheet length has not been bifurcated within the simulation domain.} %
    \label{fig:current_density}
\end{figure}

To analyze the opening up of the magnetotail in greater detail, we locate the distance from the planet where the current sheet bifurcates in the night-side region. Figure \ref{fig:current_density}(a) shows the current density plot in $xz$-plane (left) and corresponding current density magnitude ($J_{mag}$) variation along the subsolar line (right). Since the current sheet has a finite thickness along $y$ and $z$ directions, we consider a hypothetical rectangular box outside the planet's atmosphere encapsulating the current sheet and extending from -0.5 $R_{p}$ to 0.5 $R_{p}$ along both $y$ and $z$ directions. Thereafter, the spatially averaged $J_{mag}$ is calculated on $yz$ slices of the cube and its variation along $x$-direction is plotted. We find that the current sheet starts to bifurcate typically when the current density magnitude drops off to about 45$\%$ - 50$\%$ of its peak value. These are represented by line B and line A respectively in Figure \ref{fig:current_density}(a). This result is robust considering all different sets (Table \ref{table:1}) of our simulations. The current sheet, extending from the planet to its bifurcation point on the night-side region, is hereby referred to as the current sheet length.
Variation of current sheet length with planetary \& stellar wind magnetic field is shown in Figure \ref{fig:current_density}(b). As the planetary magnetic field ${B_p}$ increases, the bifurcation point recedes away from the planet. For ${B_p}$ = 2${B_e}$ we do not see any bifurcation within the simulation domain until  $|{B_{sw}}|$ = 30 nT. For a stronger planetary magnetic field of ${B_p}$ = 5${B_e}$, bifurcation starts around $|{B_{sw}}|$ = 50 nT onwards. 

It is evident from Figure \ref{fig:current_density}(b) that the current sheet bifurcation point shifts nearer to the planet when either the stellar wind magnetic field increases or the planetary magnetic field weakens.

\begin{figure}[hbtp]
    \centering
    \includegraphics[width=14cm]{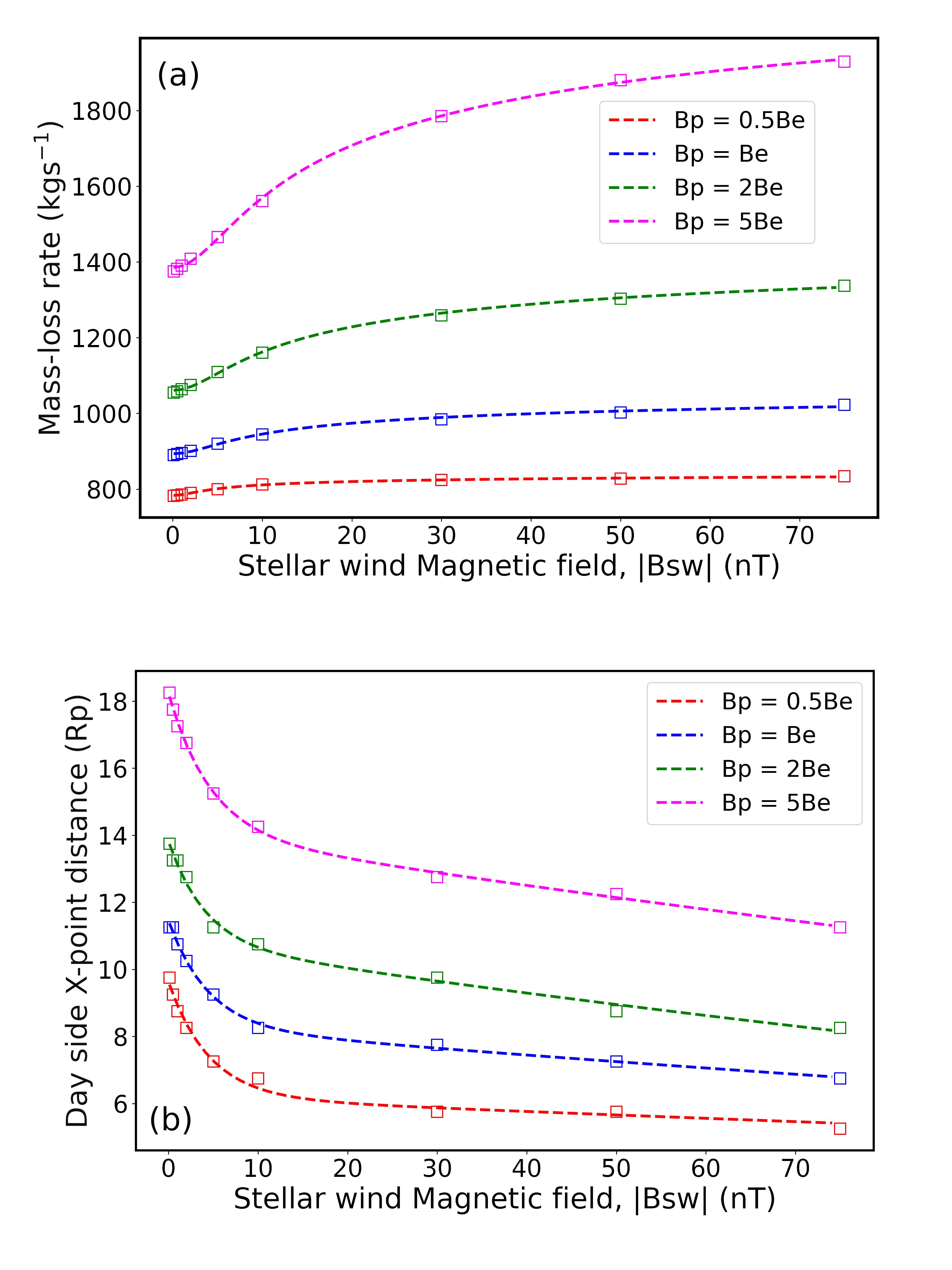}%
    \caption{Dynamics of planetary atmospheric loss: Plots of (a) total mass loss rate from a cube of side $6.6\thinspace R_p$ with the planet at its centre and (b) variation of day-side X-point distance from the planetary surface with stellar wind magnetic field strength for different magnetospheric strengths.} %
    \label{fig:massloss}
\end{figure}

We also investigate how the variations in stellar wind and planetary magnetic field strengths impact the mass-loss rate of planetary atmospheres. To compute the total mass-loss rate, we consider a cube of edge length 6.6 ${R_p}$ (extending from -3.3 ${R_p}$ to 3.3 ${R_p}$ in all three directions) with its origin coinciding with the centre of the planet such that the whole planet including its atmosphere is enclosed within the cube. Mass loss rates are computed across all six faces of the cube which contribute to the total mass loss. The rate of atmospheric mass loss depends on the extent of stellar wind penetration into the planetary atmosphere. This is essentially denoted by a point at the dayside --- the reconnection $X$ point --- beyond which planetary magnetic field lines merge with the stellar field. If the  magnetic field strength of the stellar wind is increased gradually keeping that of the planet fixed, the $X$-point is expected to shift closer to the planet inducing greater atmospheric loss. However, the distance of the $X$-point from the planet does not decrease continually with increasing stellar field. The reconnection neutral point saturates at a certain level and then onwards, the stellar field starts accumulating at the day-side [Figure \ref{fig:magneto_stellar} (d)-(f)] similar to that of an imposed magnetosphere; at this stage mass loss rate attains a limiting value.
Figure \ref{fig:massloss} (a) confirms increase in atmospheric mass-loss rate with increasing stellar wind magnetic field strength. Figure \ref{fig:massloss} (b) shows the plot of the $X$-point distance from the planetary surface with increasing stellar wind magnetic field for different values of the planetary field. This signifies that for a weak planetary field, the $X$-point distance saturates to a steady value at lower stellar magnetic fields while for a stronger planetary field, the saturation occurs for higher values of the stellar wind field. Figure \ref{fig:massloss} (a) shows that for a weaker planetary field, the loss rate increases initially and then saturates at lower values of the stellar wind field while for the stronger planetary field, the loss rate requires higher values of stellar wind field to attain saturation. The atmospheric loss rate is found to have a strong negative correlation with the X-point distance from the planet. The Pearson correlation coefficient values are -0.9800, -0.9683, -0.9753 and -0.9659 for planetary magnetic fields $0.5 B_e$, $1 B_e$, $2 B_e$, and $5 B_e$ respectively.

Figure \ref{fig:massloss}(a) also indicates that for a conducting plasma atmosphere, as considered in the present study, the mass loss rate increases with increasing strength of the planetary magnetosphere. This is because a stronger magnetosphere provides a larger interaction area with the ambient medium and therefore, may lead to greater loss when it interacts with the stellar wind. Moreover, as the atmospheric matter in our study is not considered to be charge neutral, it is expected that the atmospheric plasma will take part in magnetic reconnections and be lost in the process. Earlier studies show that the atmospheric matter stretches out and escapes through polar caps and cusps during reconnections (\citealt{IMF_effect,10.1093/mnras/stz1819,tanaka2020,10.1093/mnras/stab2947}) resulting in a larger atmospheric mass loss for a stronger magnetosphere. \cite{IMF_effect} shows that an intrinsic magnetic field does not necessarily protect a planet against atmospheric escape and the escape rate can be higher even for a highly magnetized planet. Therefore the role of intrinsic magnetic field in the protection of planetary atmosphere is a complex question.

\subsection*{Semi-Analytical Expression for Planetary Mass Loss Rates}

\begin{figure}[hbtp]
    \centering
    \includegraphics[scale= 0.5]{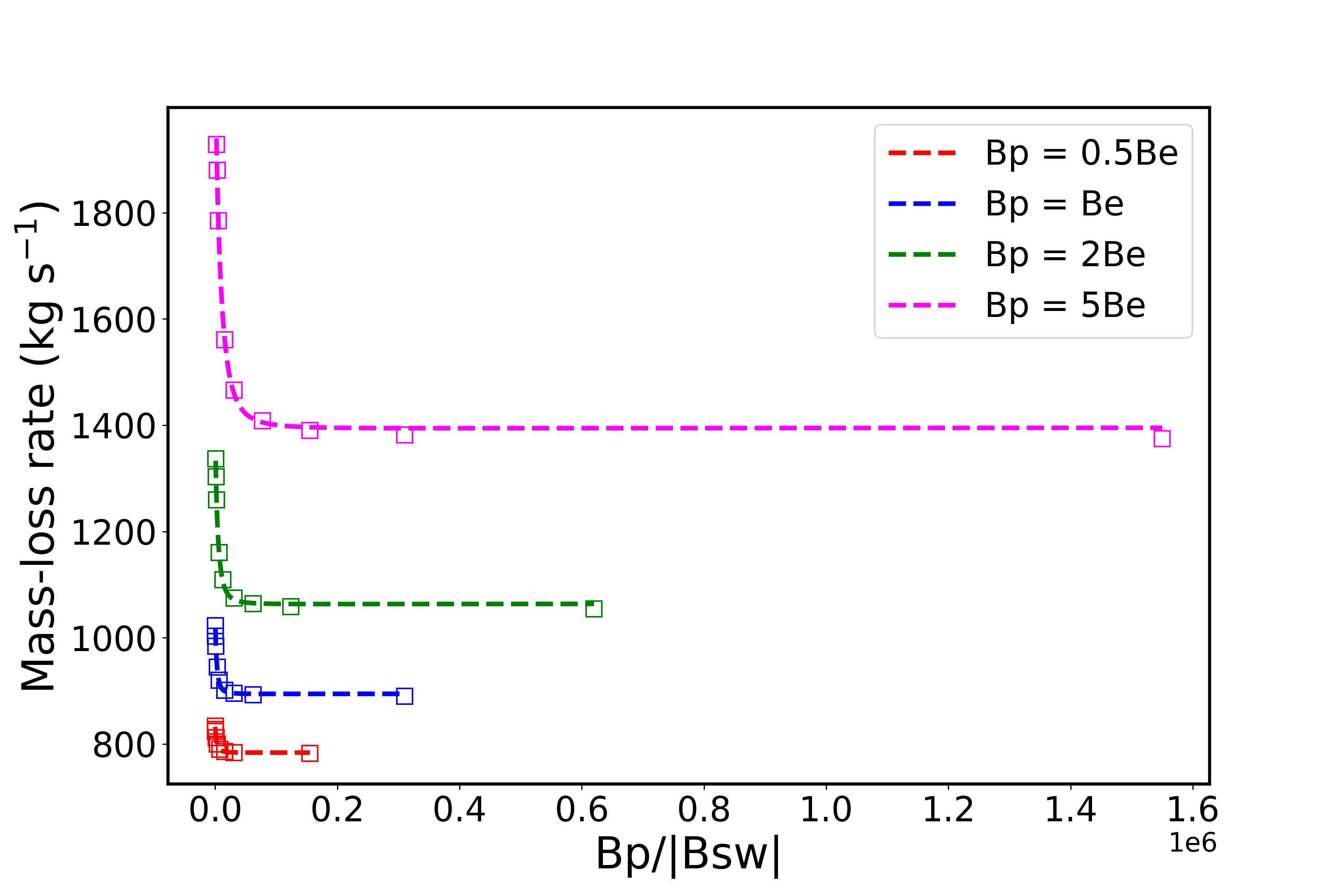}%
    \caption{Dependence of atmospheric mass loss rate on the ratio of the planetary to the stellar wind magnetic field strength. Simulation results are represented by square boxes, while the fit from the analytic expression (equation \ref{eq.13}) is shown by dotted lines.}
     \label{fig:Bp_to_Bwind_with masslossrate}%
\end{figure}

Our study shows that the global dynamics of the planet, current sheet length and atmospheric mass loss rates are significantly influenced by the relative strength of the planetary and stellar wind magnetic fields.
For different sets (Table \ref{table:1}) of our simulations, Figure \ref{fig:Bp_to_Bwind_with masslossrate} shows that ratio of planetary to stellar wind magnetic fields is an important factor in governing atmospheric mass loss rates. Based on theoretical considerations we establish an analytical relationship between these two in order to ascertain the dependence of mass loss rate on the variation of planetary and stellar wind magnetic fields.

The unpeturbed planetary dipolar magnetic field falls as $\frac{1}{r^3}$, thus pressure balance at magnetopause can be written as

\begin{equation}
  p_{dyn \thinspace sw} + \frac{B_{sw}^2}{2 \mu_0} + p_{th \thinspace sw} =   \frac{k_m^2 B_p^2 R_p^6}{2 \mu_0 R_{mp}^6} + p_{th \thinspace p} \thinspace ,
  \label{eq.9}
\end{equation}

\noindent where dynamic pressure due to stellar wind, thermal pressure due to stellar wind and thermal pressure due to planetary magnetosphere at magnetopause are depicted by $p_{dyn\thinspace sw}$ and $p_{th \thinspace sw}$, $p_{th \thinspace p}$ respectively. The symbols $B_{sw}$ and $B_{p}$ represent the stellar wind and planetary magnetic field strengths respectively while
$R_{mp}$ is the magnetopause standoff distance in terms of planetary radius $R_{p}$. The factor $k_m$ is a measure of the compression of the dipolar magnetic field by the stellar wind at the magnetopause (\citealt{mead_beard_1964,tsyganenko_2005}).

\noindent From equation \ref{eq.9}, the magnetopause stand-off distance can be expressed as

\begin{equation}
    \Big(\frac{R_{mp}}{R_p}\Big) =  \Big[\frac{k_m^2(\frac{B_p}{B_{sw}})^2}{1+ \frac{C}{B_p^2}(\frac{B_p}{B_{sw}})^2}\Big]^{1/6} \thinspace ,
    \label{eq.10}
\end{equation}

\noindent where  $C = 2 \mu_0 (p_{dyn \thinspace sw} + p_{th \thinspace sw} - p_{th \thinspace p})$.
As pointed out in section 3, the atmospheric mass loss rate ($\dot{M}$) is linearly anti-correlated with the day side X-point distance ($X_{point}$) and can be expressed as $\dot{M} = - A  X_{point} + B$ where $A$ and $B$ are fitting parameters that need to be determined. Moreover, in case of an SIMF the magnetospheric stand-off distance, $R_{mp}$ is approximately equal to the day side reconnection point ($X_{point}$). Therefore, equation \ref{eq.10} can be expressed as

\begin{equation}
    \dot{M} \simeq - A  R_p\Big[\frac{(\frac{B_p}{B_{sw}})^2}{1+ \frac{C}{B_p^2}(\frac{B_p}{B_{sw}})^2}\Big]^{1/6} + B
    \label{eq.13}
\end{equation}

\noindent Equation \ref{eq.13} gives an analytical relationship between mass loss rate and the ratio of planetary and stellar wind magnetic field strengths.

Figure \ref{fig:Bp_to_Bwind_with masslossrate} shows that our modelled output of atmospheric mass loss rate fits reasonably well with the analytical expression given by \ref{eq.13}. The value of $R_{square}$ (goodness of fit) for the modelled fit is greater than 0.99 for all instances of planetary magnetic fields. 

Despite computational limitations, we have carefully chosen a grid resolution for our simulations that balances accuracy and computational feasibility. While it is important to recognize that higher grid resolutions may result in quantitative variations in the computed atmospheric mass loss rate and slight deviations in the positions of the magnetopause and magnetotail, the main essence of our study is robust to further increment in resolution. The qualitative global large-scale structures are interdependent on parameter values and the underlying physics remains valid.

\section{CONCLUSIONS} \label{sec:displaymath}
In this study, we have used 3D global magnetohydrodynamic modelling to explore the effect of varying stellar and (exo)planetary magnetic field strengths on the steady state magnetospheric configuration and atmospheric loss rates of planets. This study is important for understanding how the magnetic activity evolution of stars affects the environment and habitability of planets with different planetary magnetic field strengths. Our results are therefore applicable to a wide domain of far-out (exo)planetary systems.

Computing pressure balance at the day-side of the planet illustrates that either strengthening the stellar wind magnetic field or weakening the planetary magnetosphere results in stellar field accumulation in front of the planet similar to that of an imposed magnetosphere. In such a scenario, the magnetopause stand-off distance cannot be obtained using the conventional procedure of balancing the dynamic pressure of the incoming wind and magnetic pressure close to the planet; a more general treatment is necessary by considering the sum of both magnetic and thermal pressures. It is found that for a particular strength of planetary magnetopshere, increasing the magnetic field of stellar wind makes the bowshock shift away from the planet while the magnetopause is formed closer, implying greater wind penetration.

We find that the relative strength of planetary and stellar wind magnetic fields plays a critical role in determining the magnetic reconnection in the magnetotail region. The long extended planetary magnetotail which exists for a moderately magnetized stellar wind or strong intrinsic planetary magnetic field ceases to exist for cases when the stellar wind magnetic field becomes extremely strong (i.e. the Alfv\'{e}nic Mach number decreases) or the planetary magnetic field becomes very weak. This leads to the formation of Alfvén wings in the night-side wake region. Our findings suggest that the magnetotail begins to open up and the point of bifurcation shifts closer to the planet when either the magnetic strength of the stellar wind is increased or the intrinsic planetary magnetic field is reduced. We observe that the magnetotail current sheet starts to bifurcate when the current density magnitude drops to about 45-50 $\%$ of its maximum value.

We also find that the atmospheric mass-loss rate increases with increase in stellar magnetic field due to greater wind penetration close to the planet. The atmospheric loss rate saturates at lower values of stellar wind magnetic field for a weak planetary field. In the case of a stronger planetary field, however, a stronger stellar wind magnetic field is required for the saturation of the atmospheric loss rate. The variation of the day-side $X$-point distance from the planet with stellar wind magnetic field provides a thorough explanation for the saturation of atmospheric loss rates since the day side $X$-point distance from the planet and atmospheric loss rate are found to have a strong negative correlation. By modelling the impact of planetary magnetic field on atmospheric escape processes, we corroborate the study by \cite{IMF_effect} that the escape rate can be higher for strongly magnetized planets.

We establish an analytical relationship between the mass loss rate and the ratio of (exo)planetary and stellar wind magnetic fields. Our study shows that the global dynamics of the planet, current sheet length, and atmospheric mass loss rate are significantly influenced by the relative magnetic field strengths of the star and the planet. The results of our numerical simulation indicate that the analytical expression derived in this study succinctly captures the modeled atmospheric mass loss rates.

The detailed parameter space study presented in this paper will be helpful for identifying stellar and planetary conditions that lead to a particular magnetospheric configuration in solar and exoplanetary systems. The results also illustrate the impact of magnetic field variability on atmospheric loss rates which is relevant for understanding the habitability of (exo)planets.

\pagebreak

\section{acknowledgments}
D.N. acknowledges useful discussions with team members of Commission E4 (Impact of Magnetic Activity on Solar and Stellar Environments) of the International Astronomical Union. The authors thank Souvik Roy, Soumyaranjan Dash and Chitradeep Saha for useful discussions. S.G. acknowledges fellowship support from University Grants Commission, Goverment of India. The  development of the Star-planet Interaction module (CESSI-SPIM) along with simulation runs were carried out at the Center of Excellence in Space Sciences India (CESSI) which is funded by IISER Kolkata, Ministry of Education, Government of India.

\bibliography{spi}{}

\begin{thebibliography}{}
\expandafter\ifx\csname natexlab\endcsname\relax\def\natexlab#1{#1}\fi
\providecommand{\url}[1]{\href{#1}{#1}}
\providecommand{\dodoi}[1]{doi:~\href{http://doi.org/#1}{\nolinkurl{#1}}}
\providecommand{\doeprint}[1]{\href{http://ascl.net/#1}{\nolinkurl{http://ascl.net/#1}}}
\providecommand{\doarXiv}[1]{\href{https://arxiv.org/abs/#1}{\nolinkurl{https://arxiv.org/abs/#1}}}

\bibitem[{Alvarado-G{\'{o}}mez {et~al.}(2020)Alvarado-G{\'{o}}mez, Drake,
  Garraffo, Cohen, Poppenhaeger, Yadav, \& Moschou}]{Alvarado_G_mez_2020}
Alvarado-G{\'{o}}mez, J.~D., Drake, J.~J., Garraffo, C., {et~al.} 2020, The
  Astrophysical Journal Letters, 902, L9, \dodoi{10.3847/2041-8213/abb885}

\bibitem[{Basak \& Nandy(2021)}]{basak_nandy_2021}
Basak, A., \& Nandy, D. 2021, Monthly Notices of the Royal Astronomical
  Society, 502, 3569–3581, \dodoi{10.1093/mnras/stab225}

\bibitem[{Brun {et~al.}(2014)Brun, García, Houdek, Nandy, \&
  Pinsonneault}]{brun_2014}
Brun, A.~S., García, R.~A., Houdek, G., Nandy, D., \& Pinsonneault, M. 2014,
  Space Science Reviews, 196, 303–356, \dodoi{10.1007/s11214-014-0117-8}

\bibitem[{Carolan {et~al.}(2021)Carolan, Vidotto, Hazra,
  Villarreal D’Angelo, \& Kubyshkina}]{10.1093/mnras/stab2947}
Carolan, S., Vidotto, A.~A., Hazra, G., Villarreal D’Angelo, C., \&
  Kubyshkina, D. 2021, Monthly Notices of the Royal Astronomical Society, 508,
  6001, \dodoi{10.1093/mnras/stab2947}

\bibitem[{Chané {et~al.}(2012)Chané, Saur, Neubauer, Raeder, \&
  Poedts}]{alfven_wings}
Chané, E., Saur, J., Neubauer, F.~M., Raeder, J., \& Poedts, S. 2012, Journal
  of Geophysical Research: Space Physics, 117, \dodoi{10.1029/2012ja017628}

\bibitem[{Cohen {et~al.}(2015)Cohen, Ma, Drake, Glocer, Garraffo, Bell, \&
  Gombosi}]{Cohen_2015}
Cohen, O., Ma, Y., Drake, J.~J., {et~al.} 2015, The Astrophysical Journal, 806,
  41, \dodoi{10.1088/0004-637x/806/1/41}

\bibitem[{Cridland {et~al.}(2017)Cridland, Pudritz, Birnstiel, Cleeves, \&
  Bergin}]{10.1093/mnras/stx1069}
Cridland, A.~J., Pudritz, R.~E., Birnstiel, T., Cleeves, L.~I., \& Bergin,
  E.~A. 2017, Monthly Notices of the Royal Astronomical Society, 469, 3910,
  \dodoi{10.1093/mnras/stx1069}

\bibitem[{Das {et~al.}(2019)Das, Basak, Nandy, \& Vaidya}]{Das_2019}
Das, S.~B., Basak, A., Nandy, D., \& Vaidya, B. 2019, The Astrophysical
  Journal, 877, 80, \dodoi{10.3847/1538-4357/ab18ad}

\bibitem[{Egan {et~al.}(2019)Egan, Jarvinen, Ma, \&
  Brain}]{10.1093/mnras/stz1819}
Egan, H., Jarvinen, R., Ma, Y., \& Brain, D. 2019, Monthly Notices of the Royal
  Astronomical Society, 488, 2108, \dodoi{10.1093/mnras/stz1819}

\bibitem[{Finley {et~al.}(2018)Finley, Matt, \& See}]{Finley_2018}
Finley, A.~J., Matt, S.~P., \& See, V. 2018, The Astrophysical Journal, 864,
  125, \dodoi{10.3847/1538-4357/aad7b6}

\bibitem[{Gallet {et~al.}(2016)Gallet, Charbonnel, Amard, Brun, Palacios, \&
  Mathis}]{gallet_charbonnel_amard_brun_palacios_mathis_2016}
Gallet, F., Charbonnel, C., Amard, L., {et~al.} 2016, Astronomy and
  Astrophysics, 597, \dodoi{10.1051/0004-6361/201629034}

\bibitem[{Grie{\ss}meier(2015)}]{article2}
Grie{\ss}meier, J.-M. 2015, Detection Methods and Relevance of Exoplanetary
  Magnetic Fields,  Cham: Springer International Publishing,
  \dodoi{10.1007/978-3-319-09749-7_11}

\bibitem[{Gronoff {et~al.}(2011)Gronoff, Mertens, Lilensten, Desorgher,
  Flückiger, \& Velinov}]{titan}
Gronoff, G., Mertens, C., Lilensten, J., {et~al.} 2011, Astronomy and
  Astrophysics, 529, \dodoi{10.1051/0004-6361/201015675}

\bibitem[{Gunell {et~al.}(2018)Gunell, Maggiolo, Nilsson, Stenberg~Wieser,
  Slapak, Lindkvist, Hamrin, \& De~Keyser}]{IMF_effect}
Gunell, H., Maggiolo, R., Nilsson, H., {et~al.} 2018, Astronomy and
  Astrophysics, 614, \dodoi{10.1051/0004-6361/201832934}

\bibitem[{Harbach {et~al.}(2021)Harbach, Moschou, Garraffo, Drake,
  Alvarado-G{\'{o}}mez, Cohen, \& Fraschetti}]{Harbach_2021}
Harbach, L.~M., Moschou, S.~P., Garraffo, C., {et~al.} 2021, The Astrophysical
  Journal, 913, 130, \dodoi{10.3847/1538-4357/abf63a}

\bibitem[{Holmberg {et~al.}(2019)Holmberg, André, Garnier, Modolo, Andersson,
  Halekas, Mazelle, Steckiewicz, Génot, Fedorov, \& et~al.}]{Holmberg_2019}
Holmberg, M.~K., André, N., Garnier, P., {et~al.} 2019, Journal of Geophysical
  Research: Space Physics, 124, 8564–8589, \dodoi{10.1029/2019ja026954}

\bibitem[{J.-D.~do Nascimento {et~al.}(2016)J.-D.~do Nascimento, Vidotto,
  Petit, Folsom, Castro, Marsden, Morin, de~Mello, Meibom, Jeffers, Guinan, \&
  Ribas}]{do_Nascimento_Jr__2016}
J.-D.~do Nascimento, J., Vidotto, A.~A., Petit, P., {et~al.} 2016, The
  Astrophysical Journal, 820, L15, \dodoi{10.3847/2041-8205/820/1/l15}

\bibitem[{Khodachenko {et~al.}(2008)Khodachenko, Lammer, Lichtenegger,
  Grießmeier, Holmström, \& Ekenbäck}]{Khodachenko}
Khodachenko, M., Lammer, H., Lichtenegger, H., {et~al.} 2008, Proceedings of
  the International Astronomical Union, 4, 283 ,
  \dodoi{10.1017/S1743921309030622}

\bibitem[{Kiefer {et~al.}(2017)Kiefer, Schad, Davies, \&
  Roth}]{kiefer_schad_davies_roth_2017}
Kiefer, R., Schad, A., Davies, G., \& Roth, M. 2017, Astronomy and
  Astrophysics, 598, \dodoi{10.1051/0004-6361/201628469}

\bibitem[{Koskinen {et~al.}(2010)Koskinen, Cho, Achilleos, \&
  Aylward}]{Koskinen_2010}
Koskinen, T.~T., Cho, J. Y.-K., Achilleos, N., \& Aylward, A.~D. 2010, The
  Astrophysical Journal, 722, 178, \dodoi{10.1088/0004-637x/722/1/178}

\bibitem[{Lammer(2013)}]{book}
Lammer, H. 2013, Origin and Evolution of Planetary Atmospheres,
  \dodoi{10.1007/978-3-642-32087-3}

\bibitem[{Lammer {et~al.}(2009)Lammer, Bredehöft, Coustenis, Khodachenko,
  Kaltenegger, Grasset, Prieur, Raulin, Ehrenfreund, Yamauchi, \&
  et~al.}]{article3}
Lammer, H., Bredehöft, J.~H., Coustenis, A., {et~al.} 2009, The Astronomy and
  Astrophysics Review, 17, 181–249, \dodoi{10.1007/s00159-009-0019-z}

\bibitem[{Lammer {et~al.}(2012)Lammer, Güdel, Kulikov, Ribas, Zaqarashvili,
  Khodachenko, Kislyakova, Gröller, Odert, Leitzinger, Fichtinger, Krauss,
  Hausleitner, Holmström, Sanz-Forcada, Lichtenegger, Hanslmeier, Shematovich,
  Bisikalo, \& Fridlund}]{article}
Lammer, H., Güdel, M., Kulikov, Y., {et~al.} 2012, Earth Planets and Space,
  64, 179, \dodoi{10.5047/eps.2011.04.002}

\bibitem[{Liu {et~al.}(1999)Liu, Nagy, Groth, DeZeeuw, Gombosi, \&
  Powell}]{Liu_Powell_1999}
Liu, Y., Nagy, A.~F., Groth, C.~P., {et~al.} 1999, Geophysical Research
  Letters, 26, 2689–2692, \dodoi{10.1029/1999gl900584}

\bibitem[{Lu {et~al.}(2015)Lu, Wang, Kabin, Zhao, Liu, Zhao, \&
  Li}]{lu_wang_kabin_zhao_liu_zhao_li_2015}
Lu, J., Wang, M., Kabin, K., {et~al.} 2015, Planetary and Space Science, 106,
  108–115, \dodoi{10.1016/j.pss.2014.12.003}

\bibitem[{Lunine {et~al.}(2009)Lunine, Macintosh, \&
  Peale}]{lunine_macintosh_peale_2009}
Lunine, J.~I., Macintosh, B., \& Peale, S. 2009, Physics Today, 62, 46–51,
  \dodoi{10.1063/1.3141941}

\bibitem[{Ma(2004)}]{Ma_2004}
Ma, Y. 2004, Journal of Geophysical Research, 109, \dodoi{10.1029/2003ja010367}

\bibitem[{Mayor \& Queloz(1995)}]{mayor_queloz_1995}
Mayor, M., \& Queloz, D. 1995, Nature, 378, 355–359, \dodoi{10.1038/378355a0}

\bibitem[{Mead \& Beard(1964)}]{mead_beard_1964}
Mead, G.~D., \& Beard, D.~B. 1964, Journal of Geophysical Research, 69,
  1169–1179, \dodoi{10.1029/jz069i007p01169}

\bibitem[{Mignone {et~al.}(2007)Mignone, Bodo, Massaglia, Matsakos, Tesileanu,
  Zanni, \& Ferrari}]{Mignone_2007}
Mignone, A., Bodo, G., Massaglia, S., {et~al.} 2007, The Astrophysical Journal
  Supplement Series, 170, 228, \dodoi{10.1086/513316}

\bibitem[{Nandy(2004)}]{nandy_2004}
Nandy, D. 2004, Solar Physics, 224, 161–169,
  \dodoi{10.1007/s11207-005-4990-x}

\bibitem[{Nandy \& Martens(2007)}]{NANDY2007891}
Nandy, D., \& Martens, P. 2007, Advances in Space Research, 40, 891,
  \dodoi{https://doi.org/10.1016/j.asr.2007.01.079}

\bibitem[{Nandy {et~al.}(2021)Nandy, Martens, Obridko, Dash, \&
  Georgieva}]{nandy_martens_obridko_dash_georgieva_2021}
Nandy, D., Martens, P.~C., Obridko, V., Dash, S., \& Georgieva, K. 2021,
  Progress in Earth and Planetary Science, 8,
  \dodoi{10.1186/s40645-021-00430-x}

\bibitem[{Nandy {et~al.}(2017)Nandy, Valio, \& Petit}]{nandy_valio_petit_2017}
Nandy, D., Valio, A., \& Petit, P. 2017, Living around active stars:
  Proceedings of the 328th symposium of the International Astronomical Union
  held in Maresias, Brazil, October 17-21, 2016 (Cambridge University Press)

\bibitem[{Oklop{\v{c}}i{\'{c}} {et~al.}(2020)Oklop{\v{c}}i{\'{c}}, Silva,
  Montero-Camacho, \& Hirata}]{Oklop_i__2020}
Oklop{\v{c}}i{\'{c}}, A., Silva, M., Montero-Camacho, P., \& Hirata, C.~M.
  2020, The Astrophysical Journal, 890, 88, \dodoi{10.3847/1538-4357/ab67c6}

\bibitem[{Pepe {et~al.}(2014)Pepe, Ehrenreich, \&
  Meyer}]{pepe_ehrenreich_meyer_2014}
Pepe, F., Ehrenreich, D., \& Meyer, M.~R. 2014, Nature, 513, 358–366,
  \dodoi{10.1038/nature13784}

\bibitem[{Pollack \& Yung(1980)}]{pollack}
Pollack, J., \& Yung, Y. 1980, Annual Review of Earth and Planetary Sciences,
  8, \dodoi{10.1146/annurev.ea.08.050180.002233}

\bibitem[{Ridley(2007)}]{Ridley2007AlfvnWA}
Ridley, A.~J. 2007, Annales Geophysicae, 25, 533

\bibitem[{Sakata {et~al.}(2020)Sakata, Seki, Sakai, Terada, Shinagawa, \&
  Tanaka}]{tanaka2020}
Sakata, R., Seki, K., Sakai, S., {et~al.} 2020, Journal of Geophysical
  Research: Space Physics, 125, \dodoi{10.1029/2019ja026945}

\bibitem[{{Saur, J.} {et~al.}(2013){Saur, J.}, {Grambusch, T.}, {Duling, S.},
  {Neubauer, F. M.}, \& {Simon, S.}}]{refId0}
{Saur, J.}, {Grambusch, T.}, {Duling, S.}, {Neubauer, F. M.}, \& {Simon, S.}
  2013, A\&A, 552, A119, \dodoi{10.1051/0004-6361/201118179}

\bibitem[{Schield(1969)}]{schield_1969}
Schield, M.~A. 1969, Journal of Geophysical Research, 74, 1275–1286,
  \dodoi{10.1029/ja074i005p01275}

\bibitem[{See {et~al.}(2014)See, Jardine, Vidotto, Petit, Marsden, Jeffers, \&
  do~Nascimento}]{vidotto_2014}
See, V., Jardine, M., Vidotto, A.~A., {et~al.} 2014, Astronomy and
  Astrophysics, 570, \dodoi{10.1051/0004-6361/201424323}

\bibitem[{Shue \& Chao(2013)}]{shue_chao_2013}
Shue, J.-H., \& Chao, J.-K. 2013, Journal of Geophysical Research: Space
  Physics, 118, 3017–3026, \dodoi{10.1002/jgra.50290}

\bibitem[{Spina {et~al.}(2020)Spina, Nordlander, Casey, Bedell, D'Orazi,
  Mel{\'{e}}ndez, Karakas, Desidera, Baratella, Galarza, \&
  Casali}]{Spina_2020}
Spina, L., Nordlander, T., Casey, A.~R., {et~al.} 2020, The Astrophysical
  Journal, 895, 52, \dodoi{10.3847/1538-4357/ab8bd7}

\bibitem[{Stevenson(1983)}]{Stevenson_1983}
Stevenson, D.~J. 1983, Reports on Progress in Physics, 46, 555,
  \dodoi{10.1088/0034-4885/46/5/001}

\bibitem[{Tripathi {et~al.}(2021)Tripathi, Nandy, \& Banerjee}]{tripathi}
Tripathi, B., Nandy, D., \& Banerjee, S. 2021, Monthly Notices of the Royal
  Astronomical Society: Letters, 506, L50, \dodoi{10.1093/mnrasl/slab035}

\bibitem[{Tsyganenko(2005)}]{tsyganenko_2005}
Tsyganenko, N.~A. 2005, Journal of Geophysical Research, 110,
  \dodoi{10.1029/2004ja010798}

\bibitem[{Turnpenney {et~al.}(2020)Turnpenney, Nichols, Wynn, \&
  Jia}]{10.1093/mnras/staa824}
Turnpenney, S., Nichols, J.~D., Wynn, G.~A., \& Jia, X. 2020, Monthly Notices
  of the Royal Astronomical Society, 494, 5044, \dodoi{10.1093/mnras/staa824}

\bibitem[{Vernisse {et~al.}(2017)Vernisse, Riousset, Motschmann, \&
  Glassmeier}]{vernisse_riousset_motschmann_glassmeier_2017}
Vernisse, Y., Riousset, J., Motschmann, U., \& Glassmeier, K.-H. 2017,
  Planetary and Space Science, 137, 40–51, \dodoi{10.1016/j.pss.2016.08.012}

\bibitem[{Vidotto {et~al.}(2015)Vidotto, Donati, Jardine, See, Petit, Boisse,
  Boro~Saikia, Hebrard, Jeffers, Marsden, \& Morin}]{article1}
Vidotto, A., Donati, J., Jardine, M., {et~al.} 2015, Monthly Notices of the
  Royal Astronomical Society: Letters, 455, \dodoi{10.1093/mnrasl/slv147}

\bibitem[{Vidotto(2021)}]{vidotto_2021}
Vidotto, A.~A. 2021, Living Reviews in Solar Physics, 18,
  \dodoi{10.1007/s41116-021-00029-w}

\bibitem[{Vidotto \& Cleary(2020)}]{10.1093/mnras/staa852}
Vidotto, A.~A., \& Cleary, A. 2020, Monthly Notices of the Royal Astronomical
  Society, 494, 2417, \dodoi{10.1093/mnras/staa852}

\bibitem[{Yan {et~al.}(2019)Yan, Casasayas-Barris, Molaverdikhani,
  Alonso-Floriano, Reiners, Pallé, Henning, Mollière, Chen, Nortmann, \&
  et~al.}]{caII}
Yan, F., Casasayas-Barris, N., Molaverdikhani, K., {et~al.} 2019, Astronomy and
  Astrophysics, 632, \dodoi{10.1051/0004-6361/201936396}

\bibitem[{Zhang {et~al.}(2016)Zhang, Khurana, Kivelson, Fatemi, Holmström,
  Angelopoulos, Jia, Wan, Liu, Chen, \& et~al.}]{zhang_khurana_2016}
Zhang, H., Khurana, K.~K., Kivelson, M.~G., {et~al.} 2016, Journal of
  Geophysical Research: Space Physics, 121, \dodoi{10.1002/2016ja022360}

\end{thebibliography}
\bibliographystyle{aasjournal}

\end{document}